\documentclass[]{jfm}

\usepackage{graphicx}
\usepackage{tikz}
\DeclareRobustCommand\sampleline[1]{%
  \tikz\draw[#1] (0,0) (0,\the\dimexpr\fontdimen22\textfont2\relax)
  -- (2em,\the\dimexpr\fontdimen22\textfont2\relax);%
}
\usepackage{newtxtext}
\usepackage{newtxmath}
\usepackage{natbib}
\usepackage{fdsymbol}
\usepackage{hyperref}
\hypersetup{
    colorlinks = true,
    urlcolor   = blue,
    citecolor  = blue,
}

\newcommand{\RomanNumeralCaps}[1]

\shorttitle{Oscillations in a compressible cylinder wake}
\shortauthor{S. L. Gai, K. M. Talluru and A. K. Nandhan}

\title{Unsteady separation and oscillations in a compressible wake of a cylinder}

\author{S. L. Gai, K. M. Talluru\corresp{\email{k.talluru@unsw.edu.au}},
 \and A. K. Nandhan}

\affiliation{School of Engineering and Technology, UNSW Canberra,
Campbell, ACT 2612, Australia}

\begin{document}

\maketitle

\begin{abstract}
The paper presents a new analysis and a new interpretation of oscillations observed in the experiments of near wakes of cylinders at Mach 4 and Mach 6 and Reynolds number range 2 $\times$ 10$^4$ to 5 $\times$ 10$^5$ by \cite{schmidt2015oscillations} and \cite{thasu2022strouhal}. It is shown that the presence of such oscillations is strongly Reynolds number dependent. It is further shown that there is a threshold Reynolds number below which wake unsteadiness and oscillations do not appear. Attention is drawn to the earlier experimental investigation of supersonic wakes behind cylinders and spheres by \cite{kendall1962exp} which discusses cylinder oscillations and confirms the concept of threshold Reynolds number. Following the earlier work of \cite{goldburg1965strouhal} on hypersonic wakes of spheres and cones, a Strouhal number($St_\theta$) based on total wake momentum thickness ($\theta$) is shown to be the relevant similarity parameter for correlating the hypersonic and high supersonic wakes of spheres, cones, as well as cylinders. In this limited sense, $St_\theta$ can be said to be `universal'. A universal Strouhal number based on only one geometry (\textit{i.e.,} a circular cylinder) and over a limited Reynolds number range as proposed by \cite{schmidt2015oscillations} is much too restrictive.
\end{abstract}

\section{Introduction}\label{sec:intro} 

\begin{figure}
	\centering
	\includegraphics[width=0.9\textwidth]{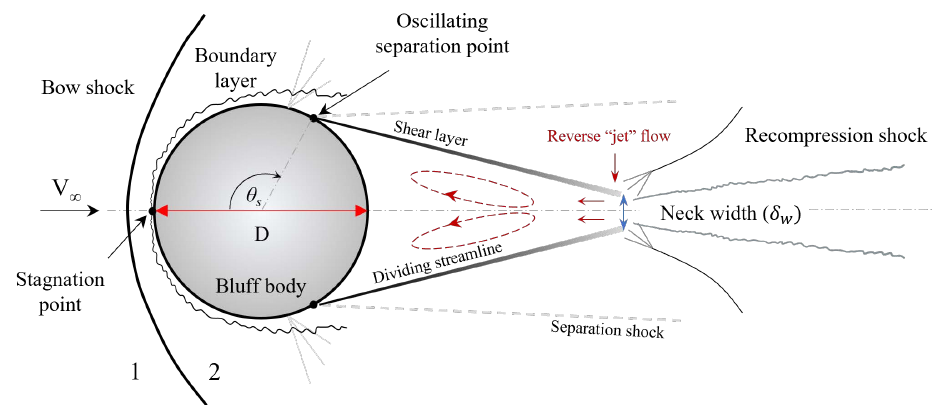}
	\caption{Schematic of the near wake region behind a circular cylinder in a supersonic/hypersonic flow.}
	\label{fig:schematic}
\end{figure}

When a blunt body is placed in a moving fluid, the flow generally separates at the rear of the body and leaves behind a wake. A typical near-wake region behind a circular cylinder at supersonic and hypersonic speeds is shown schematically in figure \ref{fig:schematic}. The freestream flow (region 1 in figure \ref{fig:schematic}) experiences strong bow shock as it encounters the bluff body in its path. In region 2 (post-shock) the flow along the axis of the body decelerates and is brought to rest at the front stagnation point at which the boundary layer originates and develops along the cylinder surface. Other features of a compressible wake are separation from the aft region of the cylinder, and the development of a free shear layer  that reattaches on the axis with a strong re-recompression  resulting in a reverse `jet' and a re-circulating flow with two counter rotating vortices bounded by the body, the shear layer, and the axis. Behind the cylinder, the two dividing streamlines extend from the separation point on the cylinder  and converge to a finite width at the `neck' of the wake. Re-compression shocks generated at the neck that turn the flow so that it is again parallel to the freestream.

Recent studies (\cite{schmidt2015oscillations}, \cite{thasu2022strouhal} and \cite{awasthi2022supersonic}) have reported sustained high frequency oscillations in the near wakes of circular cylinders exposed to supersonic and hypersonic freestream flows. It has been suggested that the origin of these oscillations is due to disturbances generated at the separation of the boundary layer on the cylinder surface and then transmitted through the resulting shear layer and fed back from the wake neck region through the subsonic part of the re-circulation region. Alternatively, disturbances originating from the neck region due to the unsteadiness of the recompression shock are transmitted upstream through the subsonic part of the shear layer and sustained by acoustic feedback mechanism of the near wake. From their experiments conducted in a Ludwieg tube at Mach 4, \cite{schmidt2015oscillations} concluded that the closed re-circulation region in the cylinder near wake acted as an acoustic resonator which sustained these oscillations. 

Based on their experiments, \cite{schmidt2015oscillations} proposed a `universal' Strouhal number of 0.48 using the dominating frequency of the power spectra measured in the wake neck region, the shear layer length (slipline) and the freestream velocity ($V_\infty$). This Strouhal number was found to be independent of Reynolds number based on freestream conditions and the cylinder diameter ($D$). The Reynolds number range of their experiments was  $2 \times 10^4$ to $ 5\times 10^5$. This work was followed by similar experiments with a cylinder at Mach 6 and Reynolds number range  $2.3 \times 10^5$ to $5 \times 10^5$ by \cite{thasu2022strouhal}. They concluded that the universal Strouhal number as defined by Schmidt $\&$ Shepherd was not only independent of Reynolds number but also Mach number.

A detailed study of cylinder wake dynamics at Mach 3 by \cite{awasthi2022supersonic} generally confirmed the acoustic resonator behaviour proposed by \cite{schmidt2015oscillations} but did not specifically confirm the universality of Strouhal number value of 0.48. In fact, based on their data, the Strouhal number based on slipline lengths at the two Reynolds numbers of their experiments are 0.51 and 0.42, respectively. It should be pointed out, however, that according to authors of \cite{awasthi2022supersonic}, the wake was determined to be turbulent at the two Reynolds numbers of $6 \times 10^5$ to $7.5 \times 10^5$ of their experiments. It should also be mentioned here that both \cite{schmidt2015oscillations} and \cite{thasu2022strouhal} do not explicitly state that the near wake was laminar or turbulent in their experiments, but we have assumed it to be laminar based on the Reynolds number range \cite[see, for example,][]{bashkin1998initiation, bashkin2002comparison}. However, this aspect is important and is discussed elaborately in the paper.

Further, \cite{schmidt2015oscillations} and \cite{thasu2022strouhal} formulated Strouhal number in terms of freestream conditions. However, looking at the schematic in figure \ref{fig:schematic}, it seems more reasonable to use the post-shock parameters for scaling the oscillations in the separated flow. This is because the boundary layer originates at the stagnation point (see the schematic in figure \ref{fig:schematic}) and develops in a region that is fully embedded in the post-shock region. 

In the light of these observations, and in view of somewhat limited range of the data, a further analysis of scaling parameters that characterise such flows, as well as new interpretations and flow features are presented here. The paper also draws some interesting comparisons with incompressible flow cylinder data and draws attention to the underlying similarities. The similarities between incompressible and compressible wakes observed in this study have further supported the use of post-shock parameters for scaling the oscillations in the wake. 

As this paper was nearing completion, we became aware of an experimental investigation of supersonic wakes behind cylinders and spheres by \cite{kendall1962exp} whose results confirm our analysis and supports the concept of threshold Reynolds number. A detailed discussion of \cite{kendall1962exp}'s data is given in the paper.
 
\section{Analysis and discussion}\label{sec:analysis}
The Strouhal number ($St_s$) proposed by \cite{schmidt2015oscillations} is defined in terms of freestream velocity and the shear layer length, measured from the separation point to the neck region (presumably up to the minimum width location). They state that the use of this length scale was motivated by studying these oscillations from shadowgraph images and also based on their Euler (inviscid) computations. However, since the shear layer oscillates, measuring the shear layer length from shadowgraph visualisation would lead to uncertainties and hence the large scatter seen in their measured data (up to $\pm\,40\,\%$). They also justify using the freestream velocity stating that freestream Reynolds numbers are high enough that inviscid regions of the flow are sufficiently independent of Reynolds number. With these assumptions, their `universal' Strouhal number is shown to be independent of Reynolds number. This was later corroborated by \cite{thasu2022strouhal}, at Mach number 6, as mentioned earlier. \cite{thasu2022strouhal} measured the shear layer length based on their schlieren flow visualisation with somewhat less scatter than those by \cite{schmidt2015oscillations}. With all these assumptions and uncertainties, the universality of this Strouhal number seems quite restrictive in the sense that it is not shown to be independent of the shape of the body producing the wake such as the universal Strouhal number as proposed by \cite{roshko1961experiments} or \cite{bearman1967vortex} for incompressible flows. It seems, therefore, still worthwhile to seek a better scaling parameter that characterises these oscillations. This is one of the aims of this paper.

\subsection{Similarity of incompressible and hypersonic (and supersonic) wakes}
\cite{fay1963unsteady} first examined unsteady hypersonic wakes behind spheres in a ballistic range and noted a close analogy with incompressible wakes. Using the Roshko number $Ro$, which is independent of velocity, they found a linear relationship between $Ro$ and $Re_h$. Here, $Re_h$ is the Reynolds number evaluated at 90 degrees measured behind the front stagnation point to represent the flow separation point. The Roshko number $Ro$ is the product of the Reynolds number $Re_h$ and the Strouhal number, where the Strouhal number $St$ ($f D /V$) is defined in the usual way.  It was found that the hypersonic wake and incompressible wake data were similar differing only in the value of linear slopes and zero intercepts of $Re_h$ (the value when the Strouhal number $St$ goes to zero) when $Ro$ was plotted against $Re_h$. The values of asymptotic Rayleigh Strouhal number $St^\ast$ and the zero intercept $Re_T$ values for the hypersonic wake and the incompressible wake were 0.7 $\&$ 0.23 and 3000 $\&$ 200 respectively. \cite{fay1963unsteady} remark: ``The fact that $St^\ast$ and $Re_T$ are so little different at Mach number ($M$) 15 from their values at $M \ll$ 1 is indeed remarkable. For these reasons alone it seems very likely that large-scale vortices are generated behind bluff bodies in hypersonic flow and, as in the incompressible case, are the major disturbances that initiate a turbulent growth of the wake". This statement is further reinforced when we note that \cite{hayes1959hypersonic} state, in connection with flow behind blunt bodies at hypersonic speeds, that: `` ....On blunt bodies in particular, local flow velocities are subsonic or moderately supersonic over much of the region of interest. .... the differences noted in hypersonic boundary layers are more of degree than of kind." This is also consistent with the Mach number independence principle (\cite{hayes1959hypersonic}) which is often invoked for analysing high Mach number ($M \geq$ 4) flows over blunt bodies including wakes.

\subsection{\texorpdfstring{Wake momentum thickness, $\theta$}%
{Wake momentum thickness}}
\cite{fay1963unsteady} point out that while the drag in incompressible flow is mainly due to the momentum defect caused by vortex shedding, the drag in hypersonic flow is entirely due to the momentum defect caused by the strong bow shock. \cite{goldburg1965strouhal} and \cite{goldburg1966transition} further explored the similarity of hypersonic and incompressible blunt body wakes of various two and three-dimensional configurations to determine the proper scaling parameter to be used. Based on their low- speed experiments on spheres and cones (\cite{goldburg1966transition}), confirmed that total wake momentum thickness ($\theta$) could be used as a correlating parameter and then \cite{goldburg1965strouhal} extended it to spheres and cones in hypersonic flow ($M = 14$) in two ballistic ranges. The Roshko number, when plotted against a representative Reynolds number similar to that used by \cite{fay1963unsteady}, and using $\theta$ as the scaling parameter, showed a linear relationship with a zero intercept when the Strouhal number went to zero. 

Equation \ref{eqn:CD} below describes the total wake momentum thickness ($\theta$) and its relation to drag coefficient ($C_D$) \cite[see][for details]{goldburg1965strouhal} as:

\begin{equation}\label{eqn:CD}
    \rho V_\infty^2 \pi^j \theta^{j + 1} = \mathrm{drag} = \frac{1}{2} \rho V_\infty^2 C_D A,
\end{equation}

\noindent where $j$ = 0 for planar and $j$ = 1 for axisymmetric flow respectively.

Thus, for a two-dimensional cylinder, the total wake momentum thickness becomes $ {\theta}/{D}  = 0.5\,C_D$, where $D$ is the cylinder diameter and $C_D$ is cylinder drag coefficient. \cite{gerrard1965disturbance}) has discussed the merits of using total wake momentum thickness $\theta$ as a characteristic length and the Strouhal number based on $\theta$ as a universal Strouhal number as has been used by \cite{goldburg1965strouhal}. He further suggests that at higher Reynolds numbers ($Re_D \geq 30,000$), \cite{goldburg1965strouhal}'s and \cite{roshko1961experiments}'s definitions of universal non-dimensional frequency ($St_\theta$ and $St_R$, respectively) are equally justifiable. It is also interesting to note that in the present case, the ratio of \cite{roshko1961experiments}'s incompressible wake width $d_w$ and total wake momentum thickness $\theta$ is 1.3 which is close to the minimum value of 1.2 quoted by \cite{gerrard1965disturbance} as the critical Reynolds number approaches. Herein we follow the approach of \cite{goldburg1965strouhal} and use the total wake momentum thickness $\theta$ as the characteristic length for analysing high supersonic and hypersonic cylinder wakes.

\cite{gowen1953drag} measured drag of cylinders in the Mach number range 0.3 $\leq$ M $\leq$ 2.9 and Reynolds number range $5 \times 10^4 \leq Re_D \leq 10^6$, and found no significant changes in the value of the drag coefficient in the supersonic Mach number range. For high supersonic and hypersonic Mach numbers (M $\geq$ 4), the continuum flow cylinder drag coefficient is about 1.24 (\cite{koppenwallner1985drag}). Then, for a two-dimensional cylinder, the wake momentum thickness becomes $ {\theta}/{D}  = 0.5 \times 1.24 = 0.62$, where $C_D$ = 1.24 is the cylinder drag coefficient using the expression in equation \ref{eqn:CD}.
 
\cite{fay1963unsteady} and \cite{goldburg1965strouhal} use the so called the shoulder Reynolds number which is a representative Reynolds number for their correlations. This has a somewhat complicated expression involving flight (freestream) velocity, initial launch tube pressure of the ballistic range before firing, and a characteristic length. It is worth noting that the shoulder Reynolds number is evaluated at (inviscid) separation point which is taken to be at 90 degrees downstream of the front stagnation point of the sphere (behind the bow shock) when the flow is parallel to the freestream. \cite{fay1963unsteady} do not make clear as to why post-shock flow conditions are not used to evaluate the shoulder Reynolds number. On the other hand, \cite{goldburg1965strouhal} do state that for cones, the shoulder Reynolds number was evaluated using flow conditions behind the cone shock.

 Now, for blunt bodies in hypersonic flows, it is a well- established convention to use the Reynolds number based on flow conditions behind the bow shock as a similarity parameter (\cite{potter1967transitional}; \cite{lukasiewicz1973critical} and \cite{koppenwallner1985drag}). Using the normal shock relations across the bow shock, it is easy to relate the freestream Reynolds and Strouhal numbers to their respective post shock values:

\begin{equation}
    Re_{2D} = Re_{1D} \bigg(\frac{\mu_1}{\mu_2}\bigg)
\end{equation}

\begin{equation}
    St_{2} = St_{1} \bigg(\frac{V_1}{V_2}\bigg),
\end{equation}

\noindent where subscripts 1 $\&$ 2 denote conditions upstream and downstream of normal shock. $Re_D$ is the Reynolds number defined as $Re_D = {\rho V D}/{\mu}$, where $\rho$ is density, $V$ is velocity, $\mu$ is viscosity, and $D$ is a characteristic length scale such as the cylinder diameter \cite[see, for example,][]{park2010laminar, hinman2017reynolds}. Similarly, the Strouhal number is defined as $St = {f D}/{V}$, where $f$ is the dominant frequency of oscillations in the wake with $V$ and $D$ as defined above.

It should be pointed out that some uncertainty exists in evaluating viscosity $\mu$ using the data of \cite{schmidt2015oscillations} and \cite{thasu2022strouhal} especially the  freestream values because of very low static temperatures. Using the Sutherland’s law and the power law with index $n$ = 0.666 for air \cite[see][pp. 27-29]{flow1991frank} and a reference temperature of 273.15\,K, the power law overestimates the viscosity $\mu_1$ by 47$\%$ in the case of \cite{schmidt2015oscillations} and by 57$\%$ in the case of \cite{thasu2022strouhal}. This discrepancy reduces to less than 1$\%$ and 4$\%$ respectively for post shock values of viscosity $\mu_2$.  Using the Lennard-Jones potential expression from \cite{hirschfelder1949viscosity} to evaluate the viscosity did not change the results significantly. Appendix \ref{appA} shows these changes of viscosity with temperature in more detail.

\begin{figure}
	\centering
	\includegraphics[width=0.65\textwidth]{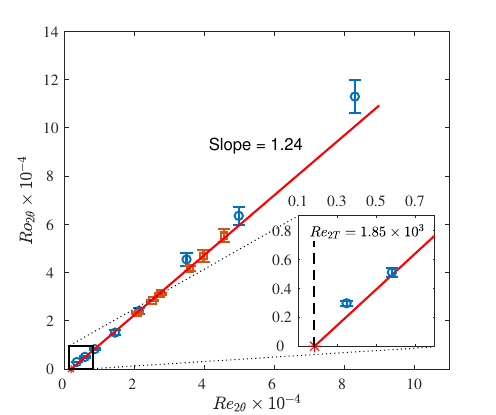}
	\caption{Roshko number ($Ro_{2 \theta}$) plotted as a function of $Re_{2 \theta}$. Symbols: $\medcircle$ - \cite{schmidt2015oscillations} and $\square$ - \cite{thasu2022strouhal}.}
	\label{fig:Ro2vsRe2}
\end{figure}

Based on the above considerations, we now re-examine the cylinder oscillations data of \cite{schmidt2015oscillations}, \cite{thasu2022strouhal}, and \cite{awasthi2022supersonic}. Figure \ref{fig:Ro2vsRe2} shows the data in terms of the Roshko number $Ro_{2 \theta} \, ( = St_{2 \theta} \times Re_{2 \theta})$ based on total wake momentum thickness ($\theta$) plotted against the post-shock Reynolds number, $Re_{2 \theta}$. The data show a linear relationship with a zero intercept of $Re_{2 T}$ = 1.85 $\times 10^3$ with a slope of 1.24, based on least square fit similar to that of \cite{goldburg1965strouhal}. The data of \cite{awasthi2022supersonic} is omitted from the figure because the near wake in their case was turbulent.

Some important inferences follow from this analysis. Firstly, it appears that the wake momentum thickness ($\theta$) and the Strouhal number based on $\theta$ can be used as similarity parameters for supersonic and hypersonic wakes. Secondly, using the Roshko number $Ro$, it is possible to characterise the threshold Reynolds number, $Re_T$, at which the wake unsteadiness first appears leading to oscillations and eventual vortex shedding. The cylinder wake data being considered here are consistent with the data from \cite{goldburg1965strouhal} for spheres and cones. Thirdly, use of both Strouhal number and Reynolds number as similarity parameters based on post-shock flow conditions for analysis of wakes of blunt bodies in high supersonic ($M_{\infty} \geq$ 4) and hypersonic flows can be justified.

\subsection{Threshold Reynolds number for oscillations}\label{sec:Threshold Re}
Next, consider the empirical relation of Rayleigh (\cite{fay1963unsteady}), which states:
\begin{equation}
    St_{2 \theta} = St^\ast \bigg[1 - \frac{Re_{2T}}{Re_{2 \theta}}\bigg],
\end{equation}
\noindent where $St^\ast$ is the asymptotic value of the Strouhal number $St_{2 \theta}$ when the Reynolds number, $Re_{2 \theta}$, is large compared to $Re_{2 T}$ which is the intercept when $St_{2 \theta}$ goes to zero. 

From figure \ref{fig:Ro2vsRe2}, the slope of the straight line is seen to be 1.24, so that
\begin{equation}
    St_{2 \theta} = 1.24 \bigg[1 - \frac{1850}{Re_{2 \theta}}\bigg].
\end{equation}

Considering further the data in figure \ref{fig:Ro2vsRe2}, we can interpret that below $Re_{2 \theta} = 1.85 \times 10^3$ (or $Re_{1D} = 1.12 \times 10^4$) oscillations will not occur as $St_{2 \theta}$ = 0 then. This would imply that there is a threshold Reynolds number below which the wake will not exhibit unsteadiness or oscillations. This may be verified when we note that most of the well-known cylinder near wake investigations done in the past at high supersonic and hypersonic flows have all been at (adiabatic or near adiabatic wall) flow conditions wherein the post-shock Reynolds numbers generally range between $10^3 \leq Re_{2D} \leq 10^4$. For example, see the data presented in Table 3 of \cite{park2010laminar}. It is also seen from the same paper (figure 12 in their paper) that in all the cylinder wake experiments (except for the low supersonic Mach number experiments of \cite{walter1953surface}, the base pressure (and hence drag) is nearly constant irrespective of Mach number and Reynolds number. It appears, therefore, that oscillations and subsequent vortex shedding is very much Reynolds number dependent as in incompressible flow. The reason, perhaps, is that hypersonic blunt body experiments quoted above were conducted at moderate to low freestream Reynolds numbers which result in stronger viscous effects in the wake (\cite{dewey1965near}) so that oscillations such as those observed by \cite{schmidt2015oscillations} or \cite{thasu2022strouhal} were not seen or were too weak to be detected. This also seems to be the case in experiments of \cite{nagata2019schlieren} wherein tests on spheres in a ballistic range at transonic and low supersonic Mach numbers in the Reynolds number range between $10^3$ and $10^5$ (based on freestream conditions and sphere diameter) showed little sign of oscillations for Reynolds numbers less than $10^5$. Further, \cite{bashkin1998initiation}, in their numerical study, also note that when $Re_{1D} \geq 10^4$, at $M = 5$, ‘flow restructuring’ and non-uniformity in the recirculating flow occurs including oscillations in velocity and temperatures as well as possibility of secondary separation. This is also in line with the observation of \cite{hinman2017reynolds} that when both Mach number and Reynolds number ranges such as those in the experiments of \cite{schmidt2015oscillations} and \cite{thasu2022strouhal} fall in the range wherein unsteadiness and eventual transition to turbulence in the near wake are possible.

These Reynolds number dependent phenomena would therefore suggest that the oscillations noted by \cite{schmidt2015oscillations} and corroborated by \cite{thasu2022strouhal} are due to the relatively high Reynolds number of their experiments and a precursor to eventual transition to turbulence and vortex shedding. At present this is only a conjecture and further experimental confirmation at both higher and lower Reynolds numbers as well as at high supersonic and hypersonic Mach numbers is definitely warranted.

\subsection{\texorpdfstring{Universality of Strouhal number, $St_\theta$}%
{University of Strouhal number}}
As discussed above, \cite{fay1963unsteady}and \cite{goldburg1965strouhal} appear to be the first to use the Roshko number and the Strouhal number to analyse the wakes of blunt bodies at hypersonic speeds. \cite{goldburg1965strouhal} then showed that the Strouhal  number, based on the total wake momentum thickness ($\theta$)  was independent of the shape of the body producing the wake. They based their conclusions on their experiments on spheres and cones. In using the Roshko number, these authors used a Reynolds number based on flight/freestream conditions. Since the Roshko number is independent of velocity, it is instructive here to analyse the \cite{schmidt2015oscillations} and \cite{thasu2022strouhal} data along with those of \cite{goldburg1965strouhal} to see if any such correlation exists for cylinder wakes.

\begin{figure}
	\centering
	\includegraphics[width=0.65\textwidth]{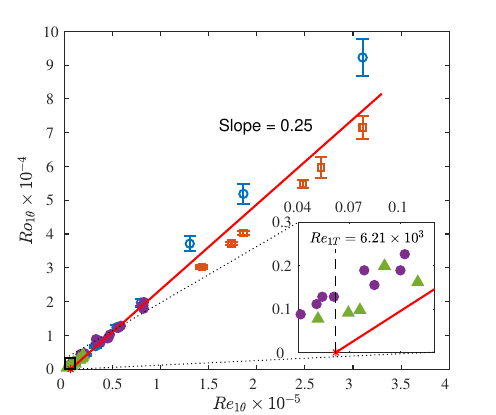}
	\caption{Roshko number ($Ro_{1 \theta}$) plotted as a function of $Re_{1 \theta}$. Symbols: $\medcircle$ - \cite{schmidt2015oscillations}; $\square$ - \cite{thasu2022strouhal}; $\medblackcircle$ - \cite[spheres,][]{goldburg1965strouhal} and $\medblacktriangleup$ - \cite[cones,][]{goldburg1965strouhal}.}
	\label{fig:Ro1vsRe1}
\end{figure}

Figure \ref{fig:Ro1vsRe1} shows these results in terms of freestream conditions unlike in figure \ref{fig:Ro2vsRe2}, where we have used the post-shock Reynolds number as similarity parameter.Since we have used Reynolds number based on freestream conditions here, it is possible to estimate the error bars for \cite{schmidt2015oscillations} and \cite{thasu2022strouhal} data as shown. It is seen that the cylinder data correlates reasonably well with the hypersonic sphere and cone data of \cite{goldburg1965strouhal}. The threshold Reynolds number $Re_{1T}$ is seen to be $6.21 \times 10^3$ and the slope is 0.25. Note that this is not too different from that of \cite{goldburg1965strouhal} value of 0.229, a difference of about 9\%. It is also noteworthy that all the data of \cite{goldburg1965strouhal} and some data of \cite{schmidt2015oscillations} are below $Re_{1 \theta} = 10^5$, while all the data of \cite{thasu2022strouhal} are at $Re_{1 \theta}  > 10^5$. We can then write:

\begin{equation}
    St_{1 \theta} = 0.25 \bigg[1 - \frac{6210}{Re_{1 \theta}}\bigg],
\end{equation}
\noindent which confirms the use of $\theta$ as the proper scaling length for the Strouhal number. 

\begin{figure}
	\centering
	\includegraphics[width=0.65\textwidth]{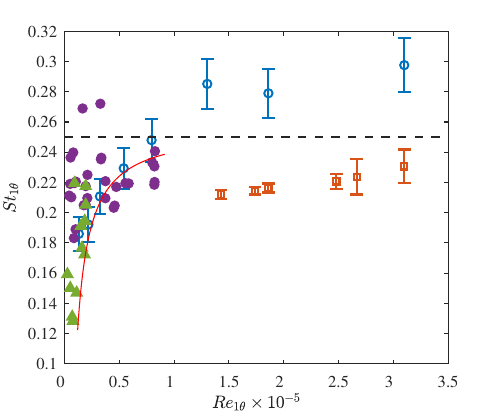}
	\caption{Strouhal number ($St_{1 \theta}$) plotted as a function of $Re_{1 \theta}$. Symbols: $\medcircle$ - \cite{schmidt2015oscillations}; $\square$ - \cite{thasu2022strouhal}; $\medblackcircle$ - \cite[spheres,][]{goldburg1965strouhal} and $\medblacktriangleup$ - \cite[cones,][]{goldburg1965strouhal}.}
	\label{fig:St1vsRe1}
\end{figure}

Further elucidation of the Strouhal number based on total wake momentum thickness $\theta$ can be seen from figure \ref{fig:St1vsRe1}, where $St_{1 \theta}$ is shown plotted against $Re_{1 \theta}$. An exponential curve is fitted through all the data points up to $Re_{1 \theta} = 10^5$. We make the following observations.  All the data of \cite{goldburg1965strouhal} are clustered close to the curve and they are all within $Re_{1 \theta} = 10^5$. There is a larger scatter in the sphere than the cone data. Note also that the lower Reynolds number data of \cite{schmidt2015oscillations} are closer to the curve as well as the sphere data. The high Reynolds number data of \cite{schmidt2015oscillations} and all the data of \cite{thasu2022strouhal} show large deviations from the fitted curve and its asymptotic value beyond $Re_{1 \theta} = 10^5$. It is therefore suggested that data at Reynolds numbers less than $Re_{1 \theta} = 10^5$ may indicate laminar (regular) oscillations, while those greater than $Re_{1 \theta} = 10^5$ may exhibit irregular oscillations. This possibility is all the more likely when we note the observations by \cite{chapman1958investigation}. 

In the context of transition in shear layers after separation from a forward-facing step geometry, \cite{chapman1958investigation} point out that at Reynolds numbers (based on the interaction length and freestream conditions) below $1\times10^5$, the separation is of the laminar type and between Reynolds numbers $1\times10^5$ and $2.5\times10^5$ it is of the transitional type. And above the Reynolds number of $2.5\times10^5$ the separation remains transitional. Since the separation here is of free interaction type, we assume that the process is at least approximately, similar in the wake in the present instance. Then, the Reynolds number based on the total momentum thickness ($\theta$) is of the same order or equivalently $2\times10^5$ based on shear layer length and freestream conditions, then half of \cite{schmidt2015oscillations} and all of \cite{thasu2022strouhal} data are greater than this (critical) Reynolds number of $Re_{1 \theta} = 10^5$, and fall into the transitional range. A further comment relating to this is given in the discussion of \cite{kendall1962exp}'s paper below. Referring to figure \ref{fig:St1vsRe1} again, an asymptotic value for the curve (shown by a dashed line) can still be identified at $St_{1 \theta} = 0.25$. It is interesting to note that this Strouhal number – Reynolds number curve resembles the low speed  $St - Re$ curve of figure 4 of \cite{roshko1954development} which delineates various flow regimes of wake behind a cylinder.

Although \cite{goldburg1965strouhal} did not claim that $St_{\theta}$ is a `universal' Strouhal number, \cite{gerrard1965disturbance} and \cite{bearman1967vortex} have labelled it as such. While \cite{gerrard1965disturbance} says that $St_{\theta}$ may qualify as a `universal' Strouhal number at high enough Reynolds number, \cite{bearman1967vortex} disagrees with this, because in an incompressible flow, $St_{\theta}$ shows dependency on the base pressure. Be that as it may, it does seem to correlate the cylinder as well as sphere and cone data reasonably well for laminar wakes at hypersonic speeds wherein the base pressure is fairly independent of both Mach and Reynolds number.

\subsection{Comparison with Kendall (1962)}

As mentioned in the Introduction, at the time this paper was nearing completion, the authors came across a very detailed experimental investigation of wakes behind cylinders and spheres at a supersonic Mach number of 3.7 and Reynolds numbers, $Re_{1D}$ ranging from $0.95 \times 10^4$ to $4.2 \times 10^4$ by \cite{kendall1962exp} conducted at the Jet Propulsion Laboratory of Caltech, USA (Technical Report No. 32-363). This investigation was conducted in a supersonic wind tunnel in which the freestream turbulence level was very low ($<$ 0.1\%). This paper is not referred to either in \cite{schmidt2015oscillations} or \cite{thasu2022strouhal}.  It is also not cited by \cite{awasthi2022supersonic}. 

In his study, \cite{kendall1962exp} conducted extensive hot wire measurements and delineated four regimes of the wakes behind cylinders based on increasing Reynolds numbers $Re_{1D}$. They are, (a) laminar steady wake; (b) laminar wake with large amplitude oscillations; (c) irregular fluctuations, which eventually break down to develop into (d) a turbulent wake. \cite{kendall1962exp} then suggests a threshold Reynolds number $Re_{1D}$ = $1.2 \times 10^4$, below which oscillations do not occur. This is the same as $Re_{1T}$ that we defined earlier and below which oscillations do not occur. It is thus consistent with and confirms our analysis. 

Interestingly, \cite{kendall1962exp} further remarks that a possible reason for the origin of these oscillations may be three-dimensional disturbances as a result of finite aspect ratio of cylinders although in his experiments, most of the time, the aspect ratio varied from 12 to 72. This is more than the aspect ratio of 10 suggested by \cite{dewey1965near} to achieve reasonably two-dimensional flow. In comparison, the aspect ratio in \cite{schmidt2015oscillations}’s experiments varied from 2.85 to 40. The aspect ratio in \cite{thasu2022strouhal}'s experiments was quite small, between 2.4 and 4. It is therefore quite possible that larger cylinder data of \cite{schmidt2015oscillations} and all of \cite{thasu2022strouhal} data are affected by spanwise flow disturbances.

Another interesting finding from \cite{kendall1962exp}'s experiments was that the oscillation frequencies varied between 50\,Hz to 850\,Hz, giving Strouhal numbers as low as 0.001 to 0.016. This appears to be based on peak fluctuation energy measured at a location of 4 diameters behind the base when the wake was still laminar. However, he comments that the fluctuation energy spectrum did not change much along the axis which he attributed to the low Reynolds number and near-stability of the wake. The wake became turbulent only at the highest Reynolds number $Re_{1D} = 4.2 \times 10^4$ of his experiments. \cite{kendall1962exp} comments that although in many ways there were similarities with low speed wakes, they could not detect frequencies with Strouhal number of 0.21. \cite{kendall1962exp} found that sphere wakes were either completely steady or completely unsteady depending on the threshold Reynolds number. At lower Reynolds numbers it was steady but became highly unsteady and three-dimensional at higher Reynolds numbers ($Re_1D >  2.5 \times 10^4$). This is clearly reflected in the relatively large scatter of sphere wake data (compared to cylinder and cone) as seen in figure \ref{fig:St1vsRe1}, , especially in the higher Reynolds numbers range. The threshold Reynolds number in the case of spheres was nearly twice that of the cylinder.
 
\subsection{\texorpdfstring{Wake neck thickness ($\delta_w$) and Reynolds number}%
{Wake neck thickness and Reynolds number}}
\begin{figure}
	\centering
	\includegraphics[width=0.65\textwidth]{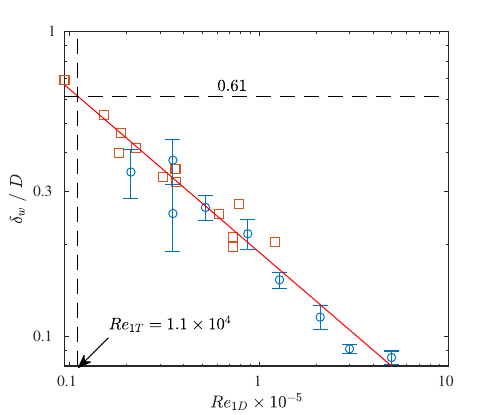}
	\caption{Wake neck thickness ($\delta_w$) plotted as a function of $Re_{1D}$. Symbols: $\medcircle$ - \cite{schmidt2015oscillations} and $\square$ - \cite{dewey1965near}.}
	\label{fig:neckthickness}
\end{figure}

As illustrated in the schematic diagram of figure \ref{fig:schematic}, the deceleration of the flow behind the cylinder and the coalescence of the shear layers shed by the separated boundary layer forms a `neck'.  The neck region at reattachment of the near wake is typical of strong viscous effect usually associated with low Reynolds number and high Mach number hypersonic flow.  This is unlike the Chapman isentropic re-compression. \cite{lees1964supersonic} and \cite{dewey1965near} have shown that with an initial finite thickness shear layer, the wake neck thickness varies with Reynolds number. Figure \ref{fig:neckthickness} shows this variation for the data of \cite{schmidt2015oscillations} along with \cite{dewey1965near}. For comparison, figure \ref{fig:neckthickness} is drawn on a log-log scale, and it is seen that approximately $\delta_w/D \sim  Re_{1D}^{-1/2}$ with  $\delta_w/D = 0.61$ at the threshold Reynolds number $Re_{1T} = 1.1 \times 10^4$. Note that this is close to the wake momentum thickness ratio $\theta/D = 0.62$. It is also interesting to note that the low Reynolds number data of \cite{schmidt2015oscillations} and \cite{dewey1965near} (which are all laminar) agree quite well below $Re_{1D} = 10^5$. 

\subsection{Further remarks on incompressible and supersonic/hypersonic cylinder wakes}
It has been known for quite some time (see, \cite{shaw1956} and Lighthill \cite{rosenhead1963laminar}) that in incompressible flow ($M$ < 1) behind a cylinder, the eddies shed into the wake and their frequency is associated with acoustic disturbances generated at separation points. This hypothesis is also supported by the experiments of \cite{schmidt2015oscillations} as discussed earlier. \cite{wu2004experimental} in their study of flow past a cylinder in low Reynolds number incompressible flow note that separation point is not stationary but fluctuates up to an angular displacement of about 3$^\circ$. \cite{fage1929xxviii} provided a remarkably detailed account of the process of separation of flow over a large cylinder at low speeds in the Reynolds number range $1 \times 10^5$ to $3.5 \times 10^5$. Further, \cite{fage1929xxviii} showed that, depending on the Reynolds number, the separation process took place over 7.5$^\circ$ and 9$^\circ$ of angular displacement at the lowest and highest Reynolds numbers, respectively. In other words, the separation took place between 76.5$^\circ$ and 84$^\circ$ at the lowest Reynolds number and between 96$^\circ$ and 105$^\circ$ at the highest Reynolds number, where the angular displacement is measured from the front stagnation point. \cite{fage1929xxviii} further showed that the separation process begins at the first inflection point (just after the pressure minimum) and is completed by the second inflection point and during which the pressure is near constant. This is followed by the separating shear layer forming a vortex sheet and growing by entraining the outer fluid into the dead air region. It is during this process that acoustic pulses are generated which are then propagated downstream resulting in oscillating flow and eddies. 

\cite{schmidt2015oscillations}, however, did not observe any unsteadiness in or movement of separation points in their experiments although their inviscid simulations did show movement of separation points over the surface. Secondly, in their experiments, the separation point varied considerably, from 105$^\circ$ presumably at the lowest Reynolds number ($2 \times 10^4$) to 120$^\circ$, at the highest Reynolds number  ($5 \times 10^5$), which does indicate some Reynolds number effect. This, however, is not made clear. In contrast, \cite{bashkin1998initiation}'s numerical simulations showed only small Reynolds and Mach number effect. This angular variation of separation point in \cite{schmidt2015oscillations} experiments with Reynolds number may be similar to what was observed by \cite{fage1929xxviii} in his low speed cylinder experiments. Further, for all the data of \cite{schmidt2015oscillations}, \cite{thasu2022strouhal}, and \cite{awasthi2022supersonic}, the \cite{bashkin1998initiation} empirical relation (based on their numerical simulations) predicts a near constant separation angle of 131 deg, while the actual separation angle varied between 8\% to 20\% (\cite{schmidt2015oscillations}), 9\% (\cite{thasu2022strouhal}), and 5\% (\cite{awasthi2022supersonic}) (see Table 2).

\section{Summary and Conclusions}

The proposal of a `universal' Strouhal number put forward by \cite{schmidt2015oscillations} has several shortcomings. Firstly, it is based on one geometry (\textit{i.e.}, circular cylinder), one Mach number, and based on a limited range of Reynolds numbers. Secondly, it uses the shear layer length as a scaling parameter, whose determination is prone to large degree of uncertainty as evidenced by large error margins in their experiments. Its corroboration by \cite{thasu2022strouhal} suffers from all the same deficiencies.

This paper presents a completely new analysis and interpretation of the above experimental observations of oscillations in wakes behind cylinders at Mach 4 and Mach 6. Following along the lines of earlier work by \cite{fay1963unsteady} and \cite{goldburg1965strouhal} on hypersonic wakes behind spheres and cones, it has been possible to interpret the cylinder wake data at high supersonic speed (\cite{schmidt2015oscillations}) and hypersonic speed (\cite{thasu2022strouhal}). Using the Roshko number and post-shock Reynolds number as a similarity parameter and total wake momentum thickness ($\theta$) as a characteristic scaling length, it is first shown that a linear relationship exists between the Roshko number and the Reynolds number and that the zero intercept of this straight line identifies a threshold Reynolds number below which unsteadiness and oscillations do not occur. This post-shock Reynolds number ($Re_{2T}$) is found to be $1.85 \times 10^3$. This implies that the phenomenon is strongly Reynolds number dependent. It is explained as to why previous cylinder wake investigations at hypersonic speeds did not exhibit such unsteadiness and oscillations.  It is pointed out that some of the data of \cite{schmidt2015oscillations} and all the data of \cite{thasu2022strouhal} fall in the transitional range. 

Attention has been drawn to earlier high supersonic Mach number cylinder wake experiments by \cite{kendall1962exp} in a very low turbulence supersonic wind tunnel, which confirms our analysis and interpretation of threshold Reynolds number and subsequent occurrence of oscillations. Another important observation from \cite{kendall1962exp}'s experiments was that the origin of the oscillations may be to three-dimensional disturbances induced by finite aspect ratio effects. This has possibly affected some of the larger cylinder  data of \cite{schmidt2015oscillations} and all of the data of \cite{thasu2022strouhal}.

It is also shown that the Strouhal number $St_\theta$ based on total wake momentum thickness ($\theta$) along with freestream Reynolds number can be used to correlate sphere, cone, as well as the cylinder wakes. The Strouhal number so obtained is found to be 0.25 which is consistent with the previous value obtained by \cite{goldburg1965strouhal}. This information is new and lends further support for the use of Strouhal number based on wake momentum thickness as a proper scaling length. In this sense, it may be called a `universal' Strouhal number for wakes at high supersonic/hypersonic speeds but only in a very restricted way. The quest for a truly universal Strouhal number continues.

\appendix
\section{Variation of air viscosity with temperature}\label{appA}
 In this section, we look at the variation of viscosity ($\mu$) with temperature as obtained by three different models. First of these models, is from \cite{hirschfelder1949viscosity} based on Lennard-Jones potential. In this model, viscosity is fitted to temperature using a power law of the form,
\begin{figure}
	\centering
	\includegraphics[width=0.65\textwidth]{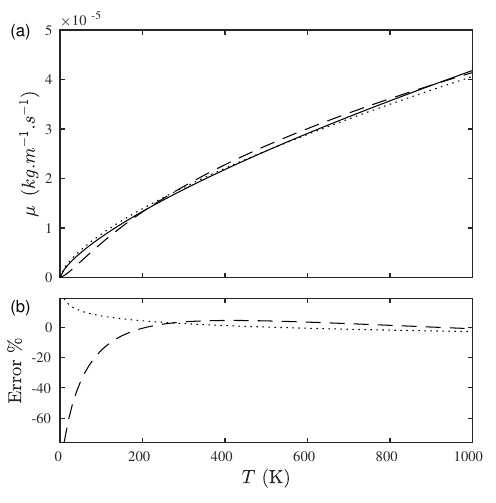}
	\caption{(a) Comparison of viscosity ($\mu$) of air obtained using Lennard-Jones potential (solid line); Sutherland's equation (dashed line) and Power law (dotted line). (b) Percentage of relative error in viscosity $\bigg( \displaystyle \frac{\mu - \mu_{_{LJ}}}{\mu_{_{LJ}}} \bigg)$ of Sutherland's equation (dashed line) and Power law (dotted line) compared to Lennard-Jones potential viscosity model, $\mu_{_{LJ}}$.}
	\label{fig:viscositymodels}
\end{figure}
 \begin{equation}
     \mu = \kappa\, T^{\omega},
 \end{equation}
\noindent where $\kappa$ = $3.1 \times 10^{-7}\, K^{-\omega}. kg. m^{-1}.s^{-1}$ and $\omega = 0.71$ were found to produce the most accurate fit. A second empirical method used for estimating the viscosity of air is the Sutherland's law (\cite{liepmann1957elements}):

\begin{equation}
    \mu = \mu_o\, \bigg(\frac{T}{T_o}\bigg)^{3/2} \,\bigg(\frac{T_o + S}{T + S}\bigg),
\end{equation}
\noindent where $\mu_o$ = $1.711 \times 10^{-5}\, kg. m^{-1}.s^{-1}$, $T_o = 273.15$ and $S = 110.4$. Lastly, the third commonly used model for viscosity is the power law based on Sutherland's constants ($\mu_o$ and $T_o$),

\begin{equation}
    \mu = \mu_o \, \bigg(\frac{T}{T_o}\bigg)^{0.666}.
\end{equation}

Figure \ref{fig:viscositymodels} shows variation of viscosity of air with temperature in the range of 0 to 1000\, K as estimated from three empirical models, namely, the Lennard-Jones potential, the Sutherland's equation and the power law. It is clearly evident that the error is greater than 10\% for $T <$ 110 \,K, beyond which all the three models provide agreeable estimates of $\mu$ for 110\,K $< T <$ 1000\,K.   

\bibliographystyle{jfm}
\bibliography{compressiblewake}

\end{document}